\documentclass[acmlarge]{acmart}

\AtBeginDocument{%
  \providecommand\BibTeX{{%
    \normalfont B\kern-0.5em{\scshape i\kern-0.25em b}\kern-0.8em\TeX}}}

\setcopyright{acmcopyright}
\copyrightyear{2020}
\acmYear{2020}

\acmJournal{IMWUT}
\acmVolume{4}
\acmNumber{4}
\acmMonth{12}

\usepackage{subcaption}

\usepackage{units}

\begin{document}

\title{Modelling Memory for Individual Re-identification in Decentralised Mobile Contact Tracing Applications}

\author{Luca Bedogni}
\authornote{Both authors contributed equally to this research.}
\email{luca.bedogni@unimore.it}
\orcid{1234-5678-9012}
\affiliation{%
  \institution{University of Modena and Reggio Emilia}
  \streetaddress{P.O. Box 1212}
  \city{Modena}
  \country{Italy}
  \postcode{43017-6221}
}

\author{Shakila Khan Rumi}
\authornotemark[1]
\email{shakilakhan.rumi@rmit.edu.au}
\affiliation{%
  \institution{RMIT University}
  \city{Melbourne}
  \state{VIC}
  \country{Australia}
  \postcode{3001}
}

\author{Flora D. Salim}
\email{flora.salim@rmit.edu.au}
\affiliation{%
  \institution{RMIT University}
  \city{Melbourne}
  \state{VIC}
  \country{Australia}
  \postcode{3001}
}


\begin{abstract}
  In 2020 the coronavirus outbreak changed the lives of people worldwide. After an initial time period in which it was unclear how to battle the virus, social distancing has been recognised globally as an effective method to mitigate the disease spread. This called for technological tools such as Mobile Contact Tracing Applications (MCTA), which are used to digitally trace contacts among people, and in case a positive case is found, people with the application installed which had been in contact will be notified. De-centralised MCTA may suffer from a novel kind of privacy attack, based on the memory of the human beings, which upon notification of the application can identify who is the positive individual responsible for the notification. Our results show that it is indeed possible to identify positive people among the group of contacts of a human being, and this is even easier when the sociability of the positive individual is low. In practice, our simulation results show that identification can be made with an accuracy of more than 90\% depending on the scenario. We also provide three mitigation strategies which can be implemented in de-centralised MCTA and analyse which of the three are more effective in limiting this novel kind of attack.  
\end{abstract}

\ccsdesc[500]{Security and privacy~Privacy protections}
\ccsdesc[500]{Security and privacy~Social aspects of security and privacy}
\ccsdesc[500]{Human-centered computing~Ubiquitous and mobile computing design and evaluation methods}

\keywords{datasets, privacy preserving, digital contact tracing, mobile computing}

\maketitle

\section{Introduction}
The coronavirus shocked the world in early 2020, when the COVID-19 illness affected millions of people, with hundred of thousands of deaths. The disease spread worldwide, forcing the World Health Organization to declare it a global pandemic on March 11th 2020~\cite{who}. The coronavirus is transmitted among individuals through air droplets which are released whenever an infected person sneezes, coughs or exhales, and can therefore infect other neighboring people. To limit the contagion, governments have issued several regulations limiting the mobility of people, to make contacts with infected people as rare as possible, hence slowing down the disease spread. Until a vaccine is registered and distributed to the people, social distancing and reducing contacts among individuals is the only effective way to contain the transmission of the virus~\cite{she20202019}. 

To gradually return to normal mobility and social interactions, governments and health operators have the need to trace the contacts of people, so that whenever an hotbed is registered, it can promptly test and eventually put into quarantine affected people. Apart from manual contact tracing, several proposals have been developed concerning Mobile Contact Tracing Applications (MCTA), which digitally trace the contacts among individuals. In case someone of the contacts is later tested as positive, a notification is sent to the previous contacts so that they are warned and can take self-isolation measures, or being tested to confirm or negate the coronavirus infection \cite{eames2003contact,ferretti2020quantifying}. Principally two technologies were envisioned: GPS has the possibility to provide data about the location of the contact and can better understand where hotbeds may be arising; the second technology is Bluetooth Low Energy (BLE), which can detect whether two devices are in close proximity, hence recording the contact between the two individuals owners of the device. The GPS is seldom used mainly due to concerns about the privacy of the users, hence BLE is the one which is is used in most countries \cite{ApplicabilityOfMCTA2020}, although with some concerns about its accuracy in determining the exact distance between two devices \cite{leith2020coronavirus}. We summarize the key differences of the different proposals in Table \ref{tab:mcta}, and we describe them more in details in Section \ref{sec:background}.

\begin{table}[t]
    \centering
    \begin{tabular}{|c|c|c|c|c|}
        \hline
        Kind & Match & Data on Server & Example countries & References \\
        \hline
        Centralized & On Server & Possibly complete data & South Korea, China & \cite{SouthKoreaMCTA} \\
        \hline
        Hybrid & On Server & Only that of positive individuals & Australia & \cite{Jhanwar2020PHyCTP} \\
        \hline
        De-centralized & On Device & Only that of positive individuals & Germany, Italy & \cite{troncoso2020decentralized,EUReportMCTA,ApplicabilityOfMCTA2020,immuni,google_exp,apple_exp} \\
        \hline
    \end{tabular}
    \caption{Different MCTA architectures}
    \label{tab:mcta}
\end{table}

Currently these applications are effectively utilized in several countries, and are helping to fight the disease spread by rapidly notifying people when in contact with a positive individual. However, there are also several concerns regarding the privacy issues which can arise due to their use \cite{shukla2020privacy}. Attackers may install cameras in places were individuals transit, along with BLE scanners. Then this data can be correlated when receiving a notification to understand who is the individual responsible for that \footnote{https://www.wired.com/story/apple-google-contact-tracing-strengths-weaknesses/}, by matching the time at which a specific code was heard together with the video surveillance. Clearly, apart from having the social contacts of people released, the possibility to identify who is the positive individual responsible for infecting other people is a serious concern which may flow into riots or violence \cite{gutirrezromero2020conflict}.

Moreover, it also exists a tradeoff between the delay the server has to wait to send the notification and the privacy issues for the end user. In fact, aggregating data from multiple users together may offer better protection against attacks, at the cost of delaying the distribution of the codes of the positive individual, hence possibly not notifying in time people which may have been infected \cite{Casella_2021Delay}.

In this work we analyze a novel kind of attack, based on the memory of the individual in de-centralised MCTA. In detail, what we show is that based on the collected codes and the memory of the individuals, it is possible to reconstruct the social trace, hence identifying the positive individuals with which someone has been in contact. This kind of attack is even easier to be performed when constrained mobility is in place, such as during lockdowns or immediately after that, since the set of people someone sees in a day is low, making it easier to identify who is the positive individual. In fact, what we also show is that the sociability level of the individual, defined as the number of unique users seen in a time window in which codes remain static, is a key parameter in understanding whether this attack can take place and to what extent. 

We show the potential and the risks associated with this kind of attack using two different datasets built with real world data, pertaining BLE traces collected in two different places. Our results indicate that this memory based attack can identify positive users based on the memory of the individuals, and it also become easier if the individual has a low sociability, such as seeing co-workers and relatives and not many other people. 

At the end of our work we also give some recommendations on how to limit the memory based identification of positive individuals, which can be summarized as: \textit{(i)} send multiple positive codes together; \textit{(ii)} send as few information as possible to notify the user, but nothing more; \textit{(iii)} adopt techniques to increase the sociability of the individuals.

The rest of this paper is structured as follows: Section \ref{sec:related} discusses related works from literature on MCTA and privacy issues; Section \ref{sec:dataset} describes the two datasets we used for our experiments; Section \ref{sec:method} presents our methodology, detailing how the memory based attack can be performed; Section \ref{sec:experiments} presents the results we obtained; Section \ref{sec:conclusion} concludes this paper, and discusses recommendations on how to better protect MCTA against these kind of attacks.
\section{Related Work}
\label{sec:related}

\subsection{Background}
\label{sec:background}
To date, three different MCTA architectures were proposed: \textit{Centralized MCTA}, in which the contacts are stored on a central server which performs the matches based on the vicinity and eventually sends the notification to the users; \textit{Hybrid MCTA}, in which data is kept on the local device of the individual, but upon being tested positive the social trace is uploaded to the server, which can notify the people who had been previously in contact; \textit{Decentralised MCTA}, in which data is kept on the local device of the individual, devices exchange random codes which change every 10-20 minutes, and upon being positive tested the codes from the previous days are sent to all the people with the application installed, and the match is eventually performed on the local device. Worldwide governments have adopted one of the aforementioned architectures, though many decided to adopt decentralised MCTA, as it is the one which seems to offer the best privacy for users \cite{EUReportMCTA,ApplicabilityOfMCTA2020} In fact, in centralised MCTA if someone has access to the data it can easily reconstruct the social interactions of any individual, with possibly dangerous privacy issues, and in Hybrid MCTA each individual is bound with a static code, hence raising again potential privacy problems.

Governments worldwide released recommendations and guidelines which help application developers and consortiums to develop and distribute MCTA. For instance the European Union released its own official document \cite{EU_guideline}, stating that these apps must be \textit{(i)} voluntarily installed; \textit{(ii)} dismantled as soon as no longer needed; \textit{(iii)} use location data only if necessary, as the purpose of MCTA is not to track people. Parallel to governments advances, also smartphone makers such as Google and Apple jointly worked together to release the Privacy Preserving Decentralized Framework inspired by the work on DP-3T \cite{troncoso2020decentralized}, to monitor contacts based on Bluetooth Low Energy (BLE) \cite{apple_exp,google_exp}. In the DP-3T framework, the BLE device broadcasts a temporary and random identifier to neighboring devices, and at the same time receives codes from other smartphones. All its own personal identifier, along with the codes received by others, are stored locally on the smartphone and never sent to any third party. Upon being tested positive, the health operator releases a code to the positive individual, which can upload to a server all its past codes to a central server, which will eventually aggregate those together and send them to all the devices with the app installed. All the data in the server is removed after 14 days \cite{troncoso2020decentralized}. There are many countries which adopted this framework, with minor adjustments, such as `Corona-Warn-App’ in Germany~\cite{corona_warn}, `Immuni’ in Italy~\cite{immuni}, `SwissCovid App’ in Switzerland~\cite{SwissCovid}. Additional applications along with their architectures and model can be found in \cite{EUReportMCTA,ApplicabilityOfMCTA2020}. 

\subsection{State of the art}
MCTA is a conventional method which has been successfully used in past to control emerging epidemic, e.g. SARS and Ebola outbreak effectively~\cite{sareen2018iot, klinkenberg2006effectiveness}. Several technologies like Wifi, BLE, GPS, and social networks were used to measure the spatial proximity between multiple users in MCTA. However, many of these MCTAs face various privacy related issues which were studied in the past and also rouse when MCTA was proposed to digitally trace contact during the COVID-19 outbreak. 

In the past, few research were devoted to ensure privacy in MCTAs for disease surveillance or outbreak monitoring. A privacy-preserving contact tracing framework named efficient privacy-preserving contact tracing for infection detection (EPIC) was proposed in~\cite{8422886}. Here, the proximity contacts were identified with hybrid wireless and localization technology and matched over encrypted contents. In~\cite{li2013privacy,zhang2013privacy}, the authors proposed security protocols to ensure privacy preserved user profile matching in mobile social networking.

During the COVID-19 pandemic, several countries have deployed MCTA nationally to control the COVID-19 transmission with limited lock-down strategies \cite{Ahmed2020SurveyCovidMCTA}. A worldwide discussion and research work started, to propose protocols which could serve this purpose but without affecting the privacy of the users.

Singapore is the first country, which deployed an MCTA called `TraceTogether' nationally to mitigate COVID-19 transmission~\cite{tracetogether}. `TraceTogether' ensures the privacy of the contacts using a protocol, BlueTrace~\cite{bay2020bluetrace}, which builds upon 4 fundamental pillars: \textit{(i)} it limits the collection of personally-identifiable
information, as the only information collected is a phone number; \textit{(ii)} de-centralised, as the data is kept on the mobile devices; \textit{(iii)} the code mobile devices send changes over time; \textit{(iv)} users have control of their data, meaning that whenever they withdraw consent, all the data stored in the health authority is deleted. However in this protocol the infected users upload their traces to server which can be accessed by health authority/government. If any attack happens in the server, the traces of the infected person can be breached in this system. Different approaches have been leveraged in~\cite{cho2020contact} to enhance the privacy in TraceTogether. They have discussed the privacy based on three different concepts: \textit{1))} privacy from snoopers; \textit{2)} privacy from contacts; and \textit{3)} privacy from authorities.

An additional privacy preserving protocol has been proposed in~\cite{bell2020tracesecure} to guarantee privacy based on the above mentioned notions. The DP-3T framework is the most popular framework among the European countries\cite{EUReportMCTA}. In this framework, the id of the infected user is broadcasted to other contacts, hence the privacy of the infected person is not preserved to his proximity contacts~\cite{troncoso2020decentralized}.

In \cite{liu2020privacy}, the authors used zero knowledge $\sum$-protocol based contact tracing app to track the contacts in privacy-preserved way in DP-3T framework. This protocol enhances the privacy of an infected user from the health authority/governments and his close contacts. Here, the authors only focused on network-based or cryptographic attack. Two different exposure notification schemes, ReBabbler and CleverParrot were introduced in ~\cite{canetti2020privacy} to ensure privacy and integrity in decentralized MCTA. The type of attack that has been addressed in this paper is adversarial attack in users' BLE chips and central server. Another privacy focused decentralized MCTA was proposed in~\cite{brack2020decentralized}, which used a distributed hash table for contact tracing. Here, two users exchange encrypted and signed message to notify about the infection. The health authority verify the infection status of a positive user publicly using a blind signature mechanism. Three different attacker models based on malicious act of health authority, infected person and not infected person were discussed and defended in this article. 

However, few of the above proposal consider this kind of attack, although not mitigating it. For instance the DP-3T framework mentions that individuals can be re-identified by other people with the help of surveillance cameras and similar tools, concluding that this kind of attack is "inherent to any proximity based notification system". In our work we present this attack in detail, and also propose some mitigation techniques which would make this kind of attack less harmful for the users.

Finally efforts were also made to understand whether BLE could actually determine the distance at which two devices stand. \cite{leith2020coronavirus} shows that BLE distance estimation is indeed a challenging task, due to considerable RSSI variations as the orientiation of the device change, or whether it is placed inside a pocket or outside. The final outcome of \cite{leith2020coronavirus} is that distance estimation only leveraging BLE alone would be hard to realize, unless other social protocols such as placing phones on tables during meeting take place. This is also similar to the proposal found in \cite{kindt2020reliable}, which shows a similar scenario and proposes instead an external wearable device, which would constrain the limits of the smartphone based solution. Nevertheless, the accuracy of the BLE is still a major concern when developing MCTA leveraging this technology.
\section{Dataset Description}
\label{sec:dataset}
We have collected Bluetooth proximity data from two different studies to model and analyze the decentralized MCTA. It includes Copenhagen Networks Study~\cite{sapiezynski2019interaction} and Social Evaluation Data~\cite{madan2011sensing}. The description of both datasets are as follows:
\subsection{Copenhagen Network Study}
Copenhagen Network Study collected Bluetooth proximity data via smartphone with other data such as the network of phone call, the network of text messages, facebook friendships and others. The purpose of this study was to have a proper understanding of social systems. The dataset was collected for 706 undergraduate students at the Technical University of Denmark over one month period in 2013. The smart phone with each student in the experiment was configured as discoverable at all time. Each device scans for nearby device in every 5 minutes. Each response contains the unique identifier of all discoverable devices and the corresponding RSSI signal.  

\subsection{Social Evolution Data}
Social Network Evolution  experiment collected Bluetooth proximity data using phone sensors with other datasets such as the network of phone call, the network of text messages, Wi-Fi connection and others. The purpose of this data collection was to study the adoption of different attributes such as political opinions, diet, exercise, obesity, eating habits, epidemiological contagion, depression and stress, dorm political issues, interpersonal relationships, and privacy by new dorm students. In this study, Bluetooth proximity data was collected for 70 undergraduate students at an undergraduate university hall between September 2008 to June 2010. Each instance of this dataset contains a sender user id, receiver user id and timestamp of the discovery. It also contains the probability of two persons being in the same floor. The probability is measured based on the RSSI signal strength. 

\subsection{Comparison}

\begin{figure*}[t]
\centering
\begin{subfigure}{0.48\textwidth}
\centering
	\includegraphics[width=\textwidth]{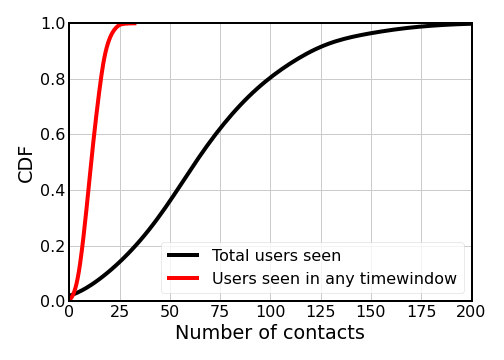}
    \caption{Copenhagen: CDF}
    \label{fig:cdf-cp}
\end{subfigure}
\quad
\begin{subfigure}{0.48\textwidth}
\centering
	\includegraphics[width=\textwidth]{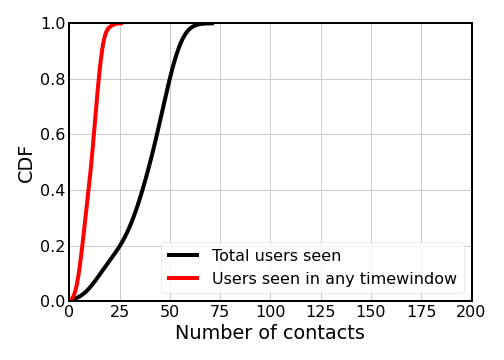}
    \caption{Social Evolution: CDF}
    \label{fig:cdf-se}
\end{subfigure}
	\caption{Cumulative distribution function for the two datasets we used in this study. Users seen in any timewindow are similar in the two datasets, while the total number of users is much bigger in the Copenhagen dataset.}
    \label{fig:cdf}
\end{figure*}

Figure \ref{fig:cdf} shows the Cumulative Distribution Function (CDF) for the two datasets with respect to the sociability of the users. In particular we plot with a red line the maximum number of users seen in any time window, while black lines refer to the total number of users seen in the whole measurement period. The maximum number of users in any timewindow highlights how crowded are the places the user visits, since it measures the number of users seen in a short time. The total number of users seen reflects instead how vast is the set of users seen during the whole measurement time, which in our case is set to 2 weeks. Although for the users seen in any timewindow both datasets show an almost identical behavior, regarding the total number of users the Social Evolution dataset shows a significantly lower number. This translates into a smaller circle of relative, or in other words people in the Social Evolution dataset tend to visit the same people more compared to the Copenhagen dataset.

These two datasets reflect well two different kind of mobility: while the Copenhagen dataset shows more users in the measurements, meaning a more diverse mobility, the Social Evolution datasets shows instead a more constrained and limited mobility, in which people is inclined to see similar people each day. With these two datasets it is then possible to study the impact of the memory based identification attack under different scenarios, which can better generalize our results. 

\section{Methodology}
\label{sec:method}

In this section we describe how we model the memory based identification attack. In our scenario, the adversary adheres to the honest-but-curious (HBC) model, in which users are participants of the communication, and do not deviate from the defined protocol. However, they will actively observe and try to understand all the possible information from the messages they receive \cite{Paverd2014ModellingAA}. In fact, in our scenario we do not look at users who crack the protocol or try to eavesdrop information not intended for them, but only look at what they purposely receive to match it with their visual encounters, eventually identifying positive and negative cases.

Let $A$ be an MCTA leveraging a decentralized protocol, hence receiving every time window $T$ codes from neighboring devices, which are then stored in $A$ along with the time at which the code is received. At the same time, user $U'$ can also keep track users who are close to her, either by writing them down or simply remembering encounters. Upon user $U''$ being tested positive, the codes are uploaded to a central platform which eventually broadcast them to all users with the application installed, including application $A$ of user $U'$. Thus $U'$ can build a graph associating all the codes to the specific time window in which it was received, along with all the users encountered in the same time window. Of course at this time it is not possible to assign a specific code to a specific individual, however with simple pruning operations in the graph it is possible to simplify it, eventually identifying positive and negative individuals.

\begin{figure*}[t]
\centering
\begin{subfigure}{0.4\textwidth}
\centering
	\includegraphics[width=\textwidth]{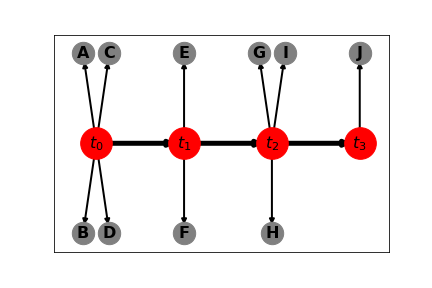}
    \caption{Step 1}
    \label{fig:step1}
\end{subfigure}
\quad
\begin{subfigure}{0.4\textwidth}
\centering
	\includegraphics[width=\textwidth]{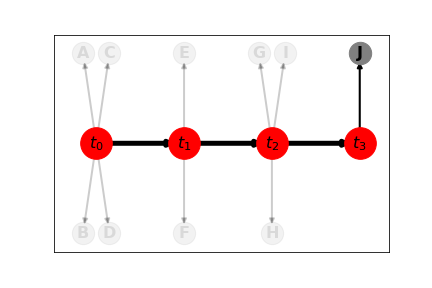}
    \caption{Step 2}
    \label{fig:step2}
\end{subfigure}
\quad
\begin{subfigure}{0.4\textwidth}
    \centering
	\includegraphics[width=\textwidth]{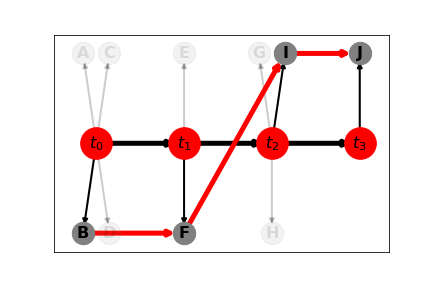}
    \caption{Step 3}
    \label{fig:step3}
\end{subfigure}
\quad
\begin{subfigure}{0.4\textwidth}
    \centering
	\includegraphics[width=\textwidth]{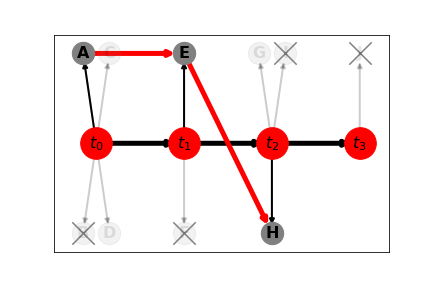}
    \caption{Step 4}
    \label{fig:step4}
\end{subfigure}
	\caption{Issues in Privacy Preserving Contact Tracing. In Figure \ref{fig:step1} we show how the BLE traces is built, as each red node represents a different time window connected to every code received in $t_0$, and to the subsequent time window. Figure \ref{fig:step2} shows that in case a single positive code \textbf{J} from a time window is received, then it is possible to uniquely identify the user who sent it. Figure \ref{fig:step3} shows instead that if multiple messages received by the user who sent \textbf{J} are received, then it is also possible to correlate them, possibly using multiple devices which share their BLE data. Finally Figure \ref{fig:step4} consequently show that this behavior may rouse a chain effect, in which also other users may be vulnerable to the identification shown in Figure \ref{fig:step2}, in this case the user sending BLE measurements \textbf{A}, \textbf{E} and \textbf{H}.}
    \label{fig:issues}
\end{figure*}

In Figure \ref{fig:issues} we outline how attacks can be performed on sequences of BLE codes recorded. The sequence of BLE measurements of user \textit{Alice} is built as Figure \ref{fig:step1} shows: $t_0$, $t_1$, $t_2$ and $t_3$ represent $4$ different time windows in which BLE codes remain the same for any user. During each of these time window, the code is univocal to each user, and it is broadcasted to neighboring devices. Hence \textit{Alice} collects codes from different users, knowing only the time at which she collected them. Moreover, she does not know whether measurements from different time windows belong to the same person or not, since BLE codes are generated randomly, hence it is not possible to directly correlate measurements recorded in any two time windows. Correspondingly, \textit{Alice} can also remember, or write down, users she sees in different time windows, which can be later correlated to the positive codes received. Clearly remembering or recording many users would be impractical, however also given the restricted mobility behavior which can be expected with such an application in place, the set of users \textit{Alice} is going to see would remain constant for a large amount of time. Hence for the sake of this example, it is safe to say that in different time windows \textit{Alice} is always together with \textit{Bob} and possibly other users. Hence, if \textit{Bob} is tested positive, his BLE codes for the past days are sent to a central server which will forward all these codes to the user which installed the MCTA. It is straightforward to note that in this example \textit{Alice} can immediately identify that \textit{Bob} is positive, since she received the \textbf{J} code and \textit{Bob} was the only person she met during $t_3$. 
However, \textit{Alice} would not receive just code \textbf{J}, but all the codes from \textit{Bob}. This is needed since the central server does not know that \textit{Alice} received code \textbf{J}, so it has to send all the measurement from user \textit{Bob}. Hence, in case \textit{Alice} receives all the messages from a single user, she can quickly correlate all those measurements together as Figure \ref{fig:step3} shows. In other words, if she receives all the positive codes sent by a single user, she just needs to correlate them to a specific individual in one time window, and she can identify \textit{Bob}. We also note that this may also happen if different mobile devices are used to collect information, which is then merged together, though this is out of the scope of this paper. A final issue is presented in Figure \ref{fig:step4}, which raises a chain reaction upon identifying and not considering codes identified upon receiving \textit{Bob} codes. Since \textit{Bob} sent codes \textbf{B}, \textbf{F}, \textbf{I} and \textbf{J}, then those codes cannot belong to any other user. Hence, in $t_1$ there is only one code left (\textbf{E}), which in case its user is tested positive will be sent again to other users. This issue is similar to that presented in Figure \ref{fig:step2}, in which only one code is received during a time window. 

To better illustrate how the memory attack identification can be performed we provide a simple example. \textit{Alice}, \textit{Bob} and \textit{Carl} meet at a certain location, and they stay there for a time equal to $2*t$ time windows in total. However, after the first time window $t$ \textit{Carl} leaves, leaving \textit{Alice} and \textit{Bob} together. Suppose that during the first $t$ window \textit{Alice} sent code \textbf{A}, \textit{Bob} sent code \textbf{B} and \textit{Carl} sent code \textbf{C}. During the subsequent time window, \textit{Alice} sent code \textbf{AA}, \textit{Bob} sent code \textbf{BB} and \textit{Carl} sent code \textbf{CC}, though it is not received by neither \textit{Alice} or \textit{Bob}. At this point, if code \textbf{C} from timewindow $t$ is received by the MCTA, no one is able to understand who has been tested positive, because it cannot be uniquely correlated to a single user. However if the notification message from the server also contains code \textbf{CC}, then \textbf{Alice} can immediately understand that the positive individual is \textit{Carl}. This can be easily done by \textit{Alice} since in the second time window she was alone with \textit{Bob} and did not received from the server the \textbf{BB} code Bob sent out during the second time window. This also exclude Bob from being positive in the first time window, hence the code she received must have been sent out by \textit{Carl}. Moreover, \textit{Bob} can also understand that \textit{Carl} was tested positive, since \textit{Bob} has not received code \textbf{AA}, hence \textit{Alice} has not been tested positive, eventually leaving only \textit{Carl} as the possible candidate for being tested positive. 

\subsection{Modelling contacts graph}
In this section we describe in the detail the model we use within this study.

Let $G=(V,E)$ be a graph with set of vertices $V$ and edges $E$. A vertex can be of 3 kinds: \texttt{T} (time window), \texttt{C} (code) or \texttt{U} (user). For each time window $t$, the user adds to $V$ all the codes received during $t$, and all the users seen. As we already stated, this can be done either by manually recording encounters, or can be modeled by leveraging memory from individuals, as we will later describe in Section \ref{sec:memory}.

Edges are built as follows:
\begin{equation}
    \forall t_i, (t_{i-1},t_i) \in E,
\end{equation}
which connects every time $t_i$ window with its predecessor $t_{i-1}$. Codes are then connected to the time window in which they have been received by the MCTA:
\begin{equation}
    \forall c_j \in t_i, (t_i,c_j) \in E,
\end{equation}
which connects every received code $c_j$ with the time window $t_i$ in which such code has been recorded. Finally, users are connected to all the codes for any time window in which that user has been seen: 
\begin{equation}
\label{eq:allconnections}
    \forall u_k,c_j \in t_i, (c_j,u_k) \in E,
\end{equation}
which connects every user $u_k$ seen in time window $t_i$ with each code $c_j$ received during such window. In this last step, we basically assign potentially each code to any user, as we do not know which user has generated which code. These edges can be pruned accordingly to a certain set of conditions, which we will describe in Section \ref{sec:graphoperations}. As a final note, the more the edges between users and codes, the more it is challenging to prune them and eventually identify positive and negative users. Again, this correlates to the sociability of the individual: in case a user sees many different people, then the recorded codes may belong to many different users; in case the set of people seen is low, fewer connections can be made, hence making it easier to uniquely identify the state of any individual.

\subsection{Graph operations}
\label{sec:graphoperations}
In this section we show 3 operations which can be performed on the graph in order to simplify it. These 3 operations are denoted as:
\begin{itemize}
    \item $I_p$: Identify positives
    \item $I_n$: Identify negatives
    \item $P$: Prune edges
\end{itemize}

The $I_p$ operation builds a set of user which are identified as positives. Whenever we are able to identify an individual $i$ as positive, it means that the user $j$ have received a positive code in every time window in which $i$ and $j$ were together. In its simplest case this can be done by looking for all the users which are connected only to positive codes. In other words, if in any \texttt{T} time window we can find a \texttt{C} node connected to only one \texttt{U} node, then such user must be positive. This operation can be further extended also to multiple users: a set of users is identified as positive if for any time window, the number of positive codes observed is equal to the total number of users seen during time window minus the number of those users which are identified has been already identified as negative. More formally this can be defined as follows: 
\begin{equation}
    \forall t, u_i \text{is positive} iff |c_p(t)| = |u(t)| - |u_n(t)|,
\end{equation}
where $c_p(t)$ is the set of positive codes at time $t$, $u(t)$ is the set of users seen at time $t$, and $u_n(t)$ is the set of negative users at time $t$.

Simply put, it checks whether the number of positive codes in a given time slot is equal to the number of users of that time slot which are already marked as negatives. If so, then all the other remaining users are marked as positives. The exact same operation can be applied also for the $I_n$ operation, by changing the count of positive users with the count of the negative users. More formally this is defined as:
\begin{equation}
    \forall t, u_i \text{is negative} iff |c_n(t)| = |u(t)| - |u_p(t)|.
\end{equation}

Identifying a user in any time window also enables to leverage such information in other time windows. For instance, in time windows in which there are two users, if one has already been identified either as positive or negative, then also the other user can be directly identified with the same methodology we just discussed and so on also for more numerous time windows. 
This is also made possible by the $P$ operation which we will discuss later in this section.

Clearly, identification is not always possible, since there may be time windows in which it is not possible to discriminate between negative and positive users. In this scenario, we leverage the $P$ operation, which simplifies the graph by pruning edges which are no longer valid considering the set of already identified positive and negative users.

The $P$ operation removes edges which are not valid anymore due to newly identified positive or negative users. It removes all the edges of a positive code connecting it to a negative user, and all the edges of a negative code connecting it to a positive user. This operation is necessary since during graph building all codes were connected to all users as Equation \ref{eq:allconnections} shown. However, after identifying positive and negative users, some of these connections may be removed, to simplify the graph and possibly leverage the identification of further users. 

These 3 steps are iterated multiple times, until no change in the graph is observe. This happens because $P$ may enable $I_p$ and $I_n$ to identify users thanks to a simpler graph, and the two identification operation allow $P$ to prune more edges.

\subsection{Modeling memory}
\label{sec:memory}
In this section we describe how our model accounts for the memory of the individuals. Clearly, any human being can remember more precisely recent events compared to those which happened days before \cite{underwood1957interference}. 

The recognition memory of individuals has been tested by several works, which show what is the probability of an individual to remember things happened in the past \cite{Recognition2007} \cite{Recognition2013}. For instance \cite{Recognition2013} reveals that users are able to recognize pictures seen one week before with an accuracy of around 90\%. \cite{Recognition2007} is closer to what we need for our work, as it also measures the ability to remember things happened 2 weeks before and in intermediate time periods. Accuracies are similar to \cite{Recognition2007}, stating a 75\% accuracy of remembering things happened 2 weeks before, 80\% at 1 week and roughly 90\% at 1 day. Regardless of the exact percentage of remembering things in the past, which may also vary from individual to individual, it has been shown that human memory is triggered by specific events, which make human being able to recognize specific things or scenarios which happened in the past \cite{tulving2002episodic} \cite{kanhabua2014triggers}.

For our work we leverage the data from \cite{Recognition2013} and use them as probabilities of remembering or not a specific user in a given time window. Given the graph $G$, we remove edges according to the probabilities themselves, so that recent days have more information compared to events more distant in time. More precisely, after having built $G$, we keep every user in any timewindow with a probability $p$, where $p$ is defined according to the following probabilities: to perform a worst case analysis, we define $p=0.75$ for events happened between two weeks and one week before receiving the notification, $p=0.80$ between 1 week and one day, and a probability $p=0.90$ for events happened in the current day \cite{Recognition2007}\cite{Recognition2013}.

We also want to note that additional tools may also been envisioned for these kind of attacks. For instance, people may also keep track of people encountered in different time windows. This can be realized by simply writing down who has been seen or not, or can even be realized by specialized apps which can help to keep track of encounters. Moreover, in places where CCTV is in place, recordings may also raise the probabilities we just presented, as it would be possible to look at encounters in specific times, if a positive code from such time window is received. Finally, already used tools such as Google Location history can be leveraged to identify places visited and exact times of visit, to better remember encounters. Hence, our analysis show to what extent the memory based identification attack can be performed when only memory from the individuals is taken into account, as other tools would raise the probabilities $p$, eventually raising the identification of users. We also note that scenario is related to remembering encounters of people rather than pictures or other things. However, we have also presented several tools which can be used to increase the ability to perform such recognition. However our model is general, since the ability for someone to remember encounters is only taken into account when simplifying the graph. A higher ability to remember things would keep more edges in the graph, improving the possibility to identify positive and negative individuals, while a worse memory would remove more graphs, possibly achieving worse results.

\subsection{Modeling mitigation}
In this section we discuss different techniques and proposal which can be used in order to mitigate the risks of a memory based individual identification. Each of these proposal will be tested in Section \ref{sec:experiments}, where we will also discuss under which circumstances they may work or not.

\subsubsection{Variable time windows}
\label{sec:method-tw}
Currently decentralised MCTA broadcast the whole set of codes from the positive user to all devices with the application installed. This bring a vast set of information to the receiving user, as it can disclose users' positivity or negativity based on the codes received and the different time windows. Let us make this example to be more clear: suppose \textit{Alice}, \textit{Bob} and \textit{Carl} are three co-workers which work in the same room, hence with a possible risk of infection due to their closeness. On a single day $d$, \textit{Carl} is tested positive, hence he is not going to work, leaving only \textit{Alice} and \textit{Bob} in the same room. If \textit{Alice} receives the notification from the MCTA, she can rapidly look through the codes received to disclose whether the responsible for the notification is \textit{Bob} or \textit{Carl}. Looking at things from \textit{Alice}'s perspective, if the responsible is \textit{Bob} then she must have received the positive codes also during day $d$ in which \textit{Carl} was not at work. On the opposite, if the responsible is \textit{Carl}, she must not have received any code on day $d$. Since she receives all the positive codes, performing this operation is simple and straightforward, and can directly identify \textit{Bob} or \textit{Carl} either as positive or negative.

At the same time, this may also work with strangers, i.e. people which are not part of the individual usual social network, but are seen seldom. This may include people met at the grocery store, or at the post office, or anyway in places in which we do not spend most of our time, hence contacts may be more diverse. In this case, receiving a wider set of codes may help the receiving user to rapidly exclude negative people seen more frequently, as codes during those time windows would not be present in the received codes. In this case, a large portion of users may be quickly pruned out of the graph, leaving only users seen less frequently which can be eventually identified.

We argue that having access to a limited set of information would eventually reduce the risk of this kind of attack. Focusing our analysis on de-centralised MCTA, this can be realized either with a push or pull logic. In case of a push logic, the server should send a subset of positive codes to the client, and wait for a feedback. In case such amount of information is sufficient to issue a notification, the no further data is required, hence the communication stops. Otherwise, the client may ask for more data, until a notification is issued or no further data is available. However, this requires for the server to know the client. In case of a pull logic, the client requests to the server whether it has data to send. In the positive case, the server sends a subset of positive codes, and the client can ask for more in case those are not sufficient to issue a notification to the user. Again, the communication stops whenever there is no more data to send, or in case the data sent is enough to show a notification to the user. Of course, both the MCTA and the server must be in the same circle of trust, in other words the server must have the guarantee that the client does not ask for further information if this is not strictly needed.

\subsubsection{Contact injection}
Another possibility to mitigate the risk of the memory based identification attack is to add codes into the report sent to the receiving users. These codes can be either false, randomly generated by the server and not belonging to any user, or they can be real, meaning that the server needs to aggregate reports from multiple positive individuals before sending them out. In the former case, the server must guarantee that the randomly generated codes do not belong to any other user. Codes are generated starting from the code used the day before. Given that these are random strings, it is unlikely that two smartphones generate the same code on the same day.

In case of adding false codes, it is a straightforward operation which can be done directly on the server. Upon receiving the $N$ codes from the user which has just been tested positive, it adds $k \cdot N$ codes which are randomly generated, with the guarantee that they cannot be correlated to any real user. Then the set of $(k+1) \cdot N$ codes is sent to all the users with the application installed, which now have $k \cdot N$ more codes to correlate to positive users, possibly generating more noise hence more difficulty in understanding who is positive or negative. Obviously the server must guarantee that the added codes cannot be related to any other individuals, otherwise it may rise the false positives number.

In case of aggregating real codes into the report, of course there is a trade off between the privacy preserving mechanism and the freshness of data to be reported. The server needs to have multiple reports from users before sending them out, hence it would not notify immediately users who have been in contact with the newly positive individual. Another option would be to send positive codes from users which have already been sent out, however these can be easily discarded by the MCTA, as it would easily see that some codes have already been received, hence it can be removed rapidly making this methodology not effective.

\subsubsection{Variable Contact Tracing Range}
A final possibility we envision to limit the memory based identification attack is to have a variable contact tracing range. This translates into having a variable threshold for the recognition of a close contact or a farther one. Obviously lower thresholds would extend the set of recorded contacts, eventually raising the number of false positives detected by the application, but with the benefit of increasing the sociability of the user, hence making the attack more challenging. This mitigation can be performed directly from the MCTA, which can adapt its own RSSI based on the sociability of the individuals. In fact, users with low sociability levels can be more prone to the memory based attack, while users which have a higher, more diverse sociability are more protected, as we will show in Section \ref{sec:experiments}.

We analyze the distribution of re-identification risk of individuals based on different memory attack models. In particular, we explore prosecutor, journalist and marketer attacker models. Prosecutor attack is a common type of attack, which re-identifies individual user based on the background information~\cite{el2012re}. In journalist attack, an individual user is re-identified based on the mapping to the `de-identified' databases which contains quasi-identifying and identifying variables~\cite{el2012re}. In marketer attack, the probability of the matching records of equivalence class is measured.

In this case, we argue that changing the RSSI threshold depending on the sociability of the individual better protects users against the above mentioned attacks. In particular, low sociability users should have a lower threshold, which eventually results in having more recorded contacts hence raising their digital sociability with respect to the app. All the needed parameters can be computed by the MCTA, which can change the RSSI threshold for any time window in which the sociability level of the individual is low. In our case we do not estimate the actual distance at which two devices are located, since we are instead interested in understanding the effect of a lower threshold, which would raise the distance at which two devices are considered in contact, and a higher threshold, which would make the MCTA more accurate in determining close contacts only, at the cost of possibly missing some of them.

\section{Experiments}
\label{sec:experiments}
In this section we present and analyze the experiments we performed on the two datasets presented in Section \ref{sec:dataset}, using the model we presented in Section \ref{sec:method} and the possible modeling mitigation techniques described too in Section \ref{sec:method}.

To perform our experiments, we build the graph $G$ for each user using the real data from the dataset, and then we run simulations to study different setups of our proposal. In any simulation, we infect one or more people from the individual social network, which eventually receives the positive codes to perform the local match and display the notification.

Since the two datasets are fundamentally different, we show the same analysis on both datasets to compare results and generalize outcomes. 

\begin{figure*}[t]
\centering
\begin{subfigure}{0.48\textwidth}
\centering
	\includegraphics[width=\textwidth]{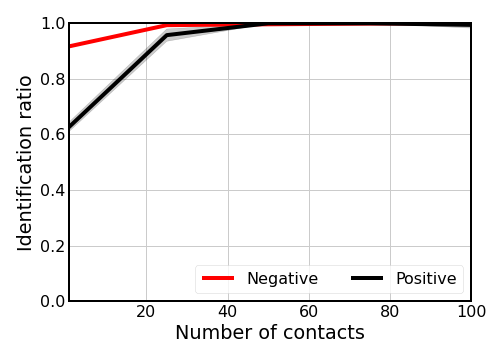}
    \caption{Copenhagen: Positive negative identification}
    \label{fig:ratio-vs-contacts-cp}
\end{subfigure}
\quad
\begin{subfigure}{0.48\textwidth}
\centering
	\includegraphics[width=\textwidth]{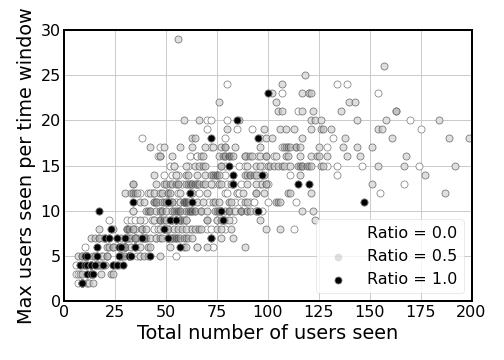}
    \caption{Copenhagen: Ratio of identification}
    \label{fig:ratio-vs-social-cp}
\end{subfigure}
	\caption{Figure \ref{fig:ratio-vs-contacts-cp} shows the identification ratio of positive and negative contacts versus the total number of users seen. Figure \ref{fig:ratio-vs-social-cp} shows the identification ratio correlating the number of users seen in a time window versus the total number of users seen during the whole measurement period. Results are obtained with the Copenhagen dataset.}
    \label{fig:res-cp}
\end{figure*}

\begin{figure*}[t]
\centering
\begin{subfigure}{0.48\textwidth}
\centering
	\includegraphics[width=\textwidth]{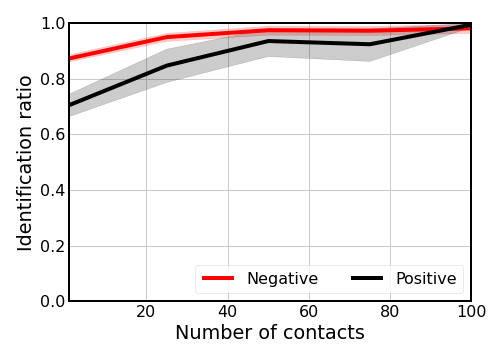}
    \caption{Social Evolution: Positive negative identification}
    \label{fig:ratio-vs-contacts-se}
\end{subfigure}
\quad
\begin{subfigure}{0.48\textwidth}
\centering
	\includegraphics[width=\textwidth]{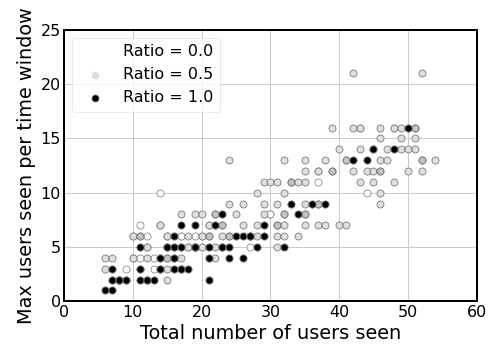}
    \caption{Social Evolution: Ratio of identification}
    \label{fig:ratio-vs-social-se}
\end{subfigure}
	\caption{Figure \ref{fig:ratio-vs-contacts-se} shows the identification ratio of positive and negative contacts versus the total number of users seen. Figure \ref{fig:ratio-vs-social-se} shows the identification ratio correlating the number of users seen in a time window versus the total number of users seen during the whole measurement period. Results are obtained with the Social Evolution dataset.}
    \label{fig:res-se}
\end{figure*}

Figure \ref{fig:res-cp} and \ref{fig:res-se} show the identification ratio of individuals based on the Copenhagen and the Social Evolution datasets, with no model mitigation in place. More precisely Figure \ref{fig:ratio-vs-contacts-cp} and Figure \ref{fig:ratio-vs-contacts-se} shows the identification ratio of positive and negative individuals versus the total number any two users had contact, for Copenhagen and Social Evolution respectively. Basically lower values on the x axis mean that two users have seldom seen each other, while higher values indicate a higher frequency of contacts. In other words, the left side of the chart shows contacts among strangers, while as we move to the right part of the chart we observe closer friends or relatives, which are seen more frequently. What Figures \ref{fig:ratio-vs-contacts-cp} and \ref{fig:ratio-vs-contacts-se} tell us is that friends or close contacts, and in general people we see more, are easier to be identified. This happens because the more we see someone, the higher the possibility to be in contact when few people is around, or to be in contact for a longer period of time. This translates into a higher chance of identifying the individual, since there are more time windows in which it is possible to perform the detection. We can see that for this analysis, the Copenhagen and Social Evolution dataset report similar results, showing that the memory based identification attack can be performed under different conditions.

Figure \ref{fig:ratio-vs-social-cp} and Figure \ref{fig:ratio-vs-social-se} show instead the ratio of identification, regardless whether of positive or negative individuals, accounting for the maximum number of users seen in any timewindow, which describe whether that user goes into crowded places or not, versus the total number of users seen during the whole measurement time. The three colors we plot which are white, gray and black, refer to identification ratios of 0, 0.5 and 1, respectively. Here we can see a clear pattern, which shows how users with low sociability levels, represented in bottom left part of the chart, tend to be far more identifiable compared to user with higher sociability levels, plotted in the top right part of the chart. This is due to the fact that having seen more users inherently adds noise into the graph, which in case it cannot be removed, makes the identification uncertain. This also confirms the hypothesis that under constrained mobility, such as during lock downs or immediately after, people will have less opportunities to meet people, hence the lower sociability level would make this kind of attack much easier to be performed. On the opposite, during normal mobility or in case for user with a higher sociability level, it would be more difficult to understand how is the responsible for the notification in case it is received, and in general it is harder to classify users seen as positive or negative. Again, also in this analysis the Copenhagen and the Social Evolution dataset show similar results.

\subsection{Variable time windows}
We now evaluate the different mitigation techniques we discussed in Section \ref{sec:method}. In this section we analyze how a variable time window to be reported to the user impacts the memory based identification attack.

\begin{figure*}[t]
\centering
\begin{subfigure}{0.48\textwidth}
\centering
	\includegraphics[width=\textwidth]{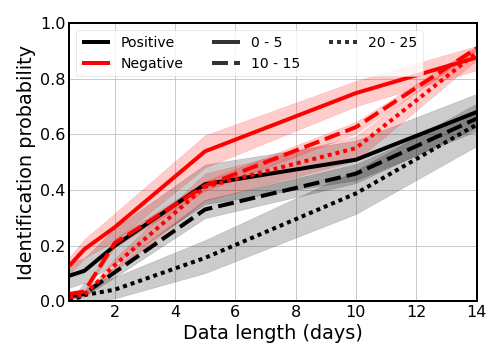}
    \caption{Copenhagen: Identification ratio versus data length}
    \label{fig:tw-cp-id-vs-l}
\end{subfigure}
\quad
\begin{subfigure}{0.48\textwidth}
\centering
	\includegraphics[width=\textwidth]{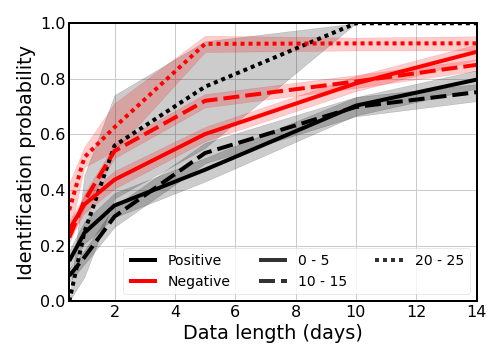}
    \caption{Social Evolution: Identification ratio versus data length}
    \label{fig:tw-se-id-vs-l}
\end{subfigure}
	\caption{Individual identification probability versus the length of the data sent to the mobile device. Results obtained with the Copenhagen datasets are shown in Figure \ref{fig:tw-cp-id-vs-l}, while results obtained with the Social Evolution dataset are plotted in Figure \ref{fig:tw-se-id-vs-l}.}
    \label{fig:tw}
\end{figure*}

Figure \ref{fig:tw-cp-id-vs-l} and Figure \ref{fig:tw-se-id-vs-l} show the identification probability of positive and negative individuals versus the length of the data sent to the mobile application by the central server, for the Copenhagen and Social Evolution datasets, respectively. The identification ratio of negative individuals is plotted in red, while the identification of positive individuals is plotted in black. Again this confirms that negative users are easier to identify compared to positive user in both datasets. Moreover, it also confirms our hypothesis formulated in Section \ref{sec:method-tw}, where we stated that the longer the set of positive codes sent to the mobile application, the easier it is to identify individuals. Again, this may seem counter intuitive, so we discuss an example to clarify it. Suppose the extreme case, in which the application receives a single positive code pertaining to a single time window in which the user saw several users. It would be impossible to identify users as positive or negative individuals based solely on such information, since such code may belong to any of the users seen in such time window. Adding a code from another window may instead help to identify users based on the codes received in the first and in the second timewindow. Again, if both codes are sensed by the MCTA, then they should belong to a positive user seen in both time windows, hence users saw only in one of the two are clearly negative. On the opposite, if the MCTA sensed only one code, then the positive user must have been seen in that specific time window, but not in the other one. The more information is added, the more it is possible to perform these operations, pruning the memory graph until an identification of an individual is made possible. This shows how reducing the set of codes sent to the application helps in mitigating the memory based identification attack.

The different line styles also show the sociability levels of individuals. For sake of readability, we only show 3 sociability levels, more precisely individuals who saw at maximum between 0 and 5 users in any timewindow (0-5), users who saw at maximum between 10 and 15 users in any timewindow (10-15), and users who saw at maximum between 20 and 25 users in any timewindow (20-25). We can clearly see that among the identification of the positive individuals, lines are ordered accordingly to the sociability level they represent, with the lower sociability (0-5) being the highest for both positive and negative identification. Higher sociability follows with lower identification probabilities. In this case the Copenhagen dataset, plotted in Figure \ref{fig:tw-cp-id-vs-l}, shows a more linear trend compared to the Social Evolution dataset shown in Figure \ref{fig:tw-se-id-vs-l}. This may happen due to the different granularity and methods of collecting data of the two datasets. Nevertheless, the general trend which shows that the more data is available the easier it is to perform the memory based identification attack holds for both.

\subsection{Contact injection}
The second mitigation technique we evaluate is the contact injection, which can be performed with false codes or real positive codes.

\begin{figure*}[t]
\centering
\begin{subfigure}{0.48\textwidth}
\centering
	\includegraphics[width=\textwidth]{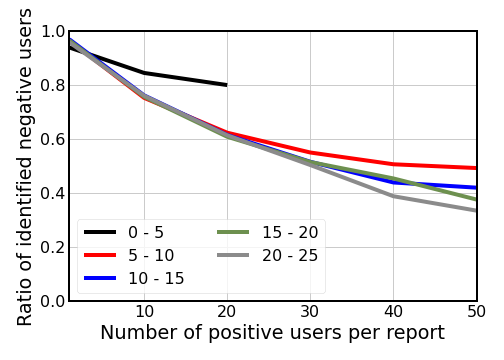}
    \caption{Copenhagen: Negative identification versus positive users per report}
    \label{fig:neg-cp-real}
\end{subfigure}
\quad
\begin{subfigure}{0.48\textwidth}
\centering
	\includegraphics[width=\textwidth]{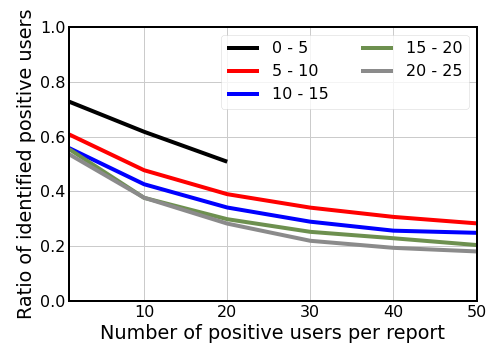}
    \caption{Copenhagen: Positive identification versus positive users per report}
    \label{fig:pos-cp-real}
\end{subfigure}
\caption{Identification of negative (Figure \ref{fig:neg-cp-real}) and positive (Figure \ref{fig:pos-cp-real}) individuals versus the number of real positive users sent by the central server. Results are obtained with the Copenhagen dataset.}
\label{fig:cp-real}
\end{figure*}

\begin{figure*}[t]
\centering
\begin{subfigure}{0.48\textwidth}
\centering
	\includegraphics[width=\textwidth]{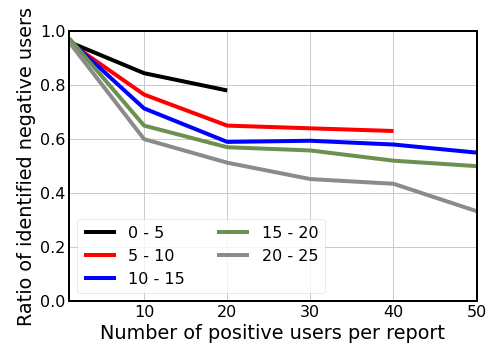}
    \caption{Social Evolution: Negative identification versus positive users per report}
    \label{fig:neg-se-real}
\end{subfigure}
\quad
\begin{subfigure}{0.48\textwidth}
\centering
	\includegraphics[width=\textwidth]{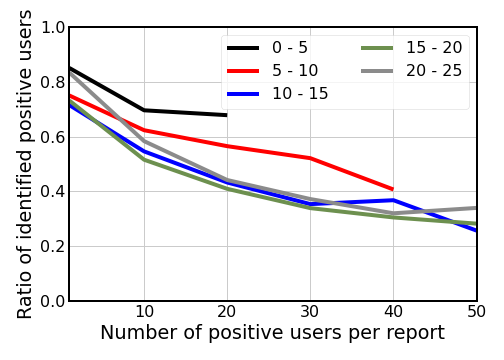}
    \caption{Social Evolution: Positive identification versus positive users per report}
    \label{fig:pos-se-real}
\end{subfigure}
\caption{Identification of negative (Figure \ref{fig:neg-cp-real}) and positive (Figure \ref{fig:pos-cp-real}) individuals versus the number of real positive users sent by the central server. Results are obtained with the Social Evolution dataset.}
\label{fig:se-real}
\end{figure*}

Figure \ref{fig:cp-real} and Figure \ref{fig:se-real} show the results when injecting real users into the report sent to the user which have the MCTA installed. A clear trend emerges, since an higher number of codes put in the report translates into a higher difficulty to identify positive and negative individuals. In fact, when adding codes from positive users into the report, the receiving user cannot easily identify the specific correlation between a user and the positive codes, hence the $I_p$ and $I_n$ operations may identify less users. This in turn makes the $P$ operation more challenging. We also note that the line for the sociability level between 0 and 5 users stops earlier, since for such users the total number of users seen in a time window was lower compared to the others, hence we evaluated less users in total. Also in this case the Copenhagen and Social Evolution dataset present similar trends, though clearer in the Copenhagen dataset, since the higher number of users in the dataset make the trend emerge better compared to the Social Evolution dataset. However both dataset clearly show that the low sociability makes the attack easier to be performed, and that by adding positive users into the dataset it is possible to mitigate the risks associated with this kind of attack.

\begin{figure*}[t]
\centering
\begin{subfigure}{0.31\textwidth}
\centering
	\includegraphics[width=\textwidth]{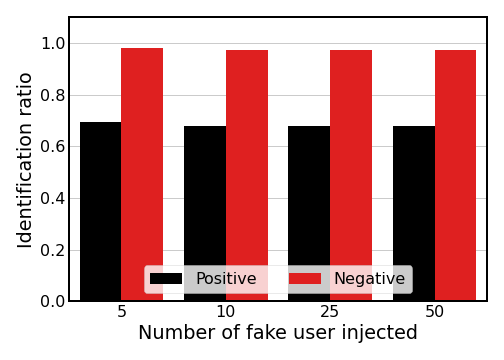}
    \caption{Copenhagen: Identification with fake injection - 1 positive user}
    \label{fig:cp-fake-1}
\end{subfigure}
\quad
\begin{subfigure}{0.31\textwidth}
\centering
	\includegraphics[width=\textwidth]{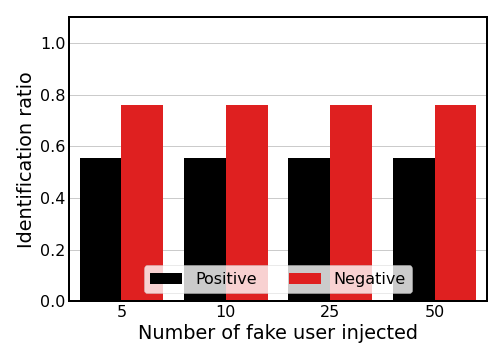}
    \caption{Copenhagen: Identification with fake injection - 10 positive users}
    \label{fig:cp-fake-10}
\end{subfigure}
\quad
\begin{subfigure}{0.31\textwidth}
\centering
	\includegraphics[width=\textwidth]{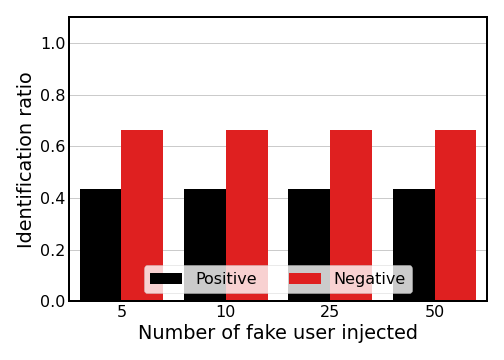}
    \caption{Copenhagen: Identification with fake injection - 20 positive users}
    \label{fig:cp-fake-20}
\end{subfigure}
\caption{Identification with injection of fake users for different number of real positive users in the report: Figure \ref{fig:cp-fake-1} presents 1 positive user, Figure \ref{fig:cp-fake-10} presents 10 positive users, and Figure \ref{fig:cp-fake-20} presents 20 positive users. Results obtained with the Copenhagen dataset.}
\label{fig:cp-fake}
\end{figure*}

\begin{figure*}[t]
\centering
\begin{subfigure}{0.31\textwidth}
\centering
	\includegraphics[width=\textwidth]{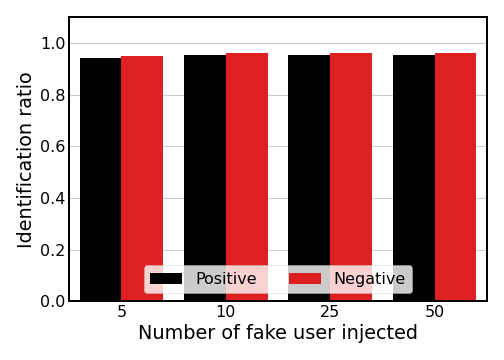}
    \caption{Social Evolution: Identification with fake injection - 1 positive user}
    \label{fig:se-fake-1}
\end{subfigure}
\quad
\begin{subfigure}{0.31\textwidth}
\centering
	\includegraphics[width=\textwidth]{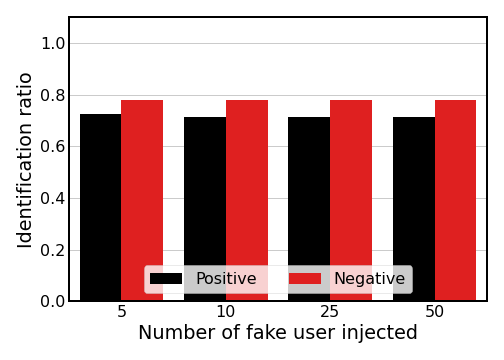}
    \caption{Social Evolution: Identification with fake injection - 10 positive users}
    \label{fig:se-fake-10}
\end{subfigure}
\quad
\begin{subfigure}{0.31\textwidth}
\centering
	\includegraphics[width=\textwidth]{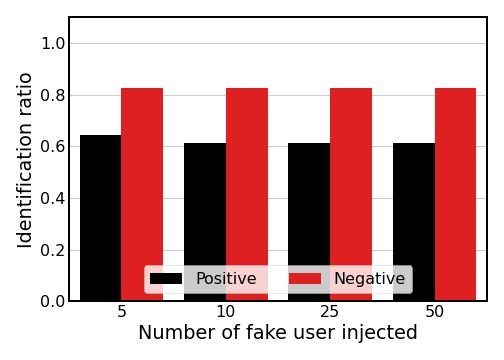}
    \caption{Social Evolution: Identification with fake injection - 20 positive users}
    \label{fig:se-fake-20}
\end{subfigure}
\caption{Identification with injection of fake users for different number of real positive users in the report: Figure \ref{fig:se-fake-1} presents 1 positive user, Figure \ref{fig:se-fake-10} presents 10 positive users, and Figure \ref{fig:se-fake-20} presents 20 positive users. Results obtained with the Social Evolution dataset.}
\label{fig:se-fake}
\end{figure*}

We now test whether the injection of false positive codes into the report helps in mitigating the memory based identification attack.

Figure \ref{fig:cp-fake} and Figure \ref{fig:se-fake} show the identification ratio when adding fake users to the report sent to the users. We show this for 1 real positive user in the notification in Figure \ref{fig:se-fake-1}, for 10 positive users added to the report in Figure \ref{fig:se-fake-10}, and for 20 positive users in Figure \ref{fig:se-fake-20}. Results confirm again that injecting codes from real positive users reduces the risk of being identified, while on the opposite injecting false positive codes does not change at all the identification ratio, regardless of the number of real positive users in the report. This can be clearly seen in any of the 3 charts of Figures \ref{fig:cp-fake}-\ref{fig:se-fake}, since no difference can be seen with respect to the identification of positive or negative individuals in the same chart. Again also in this case both datasets show a similar trend,  even though the Social Evolution datasets exhibit less differences between the identification of positive and negative individuals. 

\subsection{Variable Contact Tracing Range}
The last mitigation technique which we evaluate is changing the RSSI threshold to increase the sociability of the users. We perform this test only for the Copenhagen dataset, since the Social Evolution dataset only reports that two users are close, but no RSSI information is available. We use the API of ARX data anonymization tool~\footnote{https://arx.deidentifier.org/} to measure the re-identification risk using different privacy attack models. In the following discussion, we particularly report the re-identification risk of `prosecutor' attack model as it is the most common type of memory based attack.

\begin{figure*}[t]
\centering
\begin{subfigure}{0.48\textwidth}
\centering
	\includegraphics[width=\textwidth]{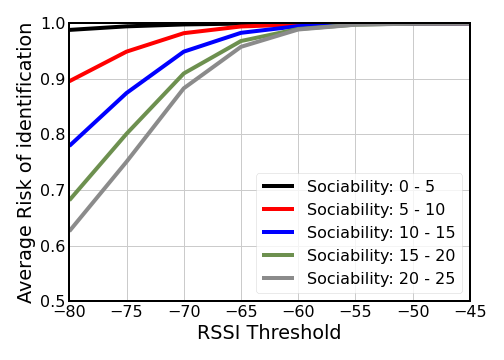}
    \caption{Identification risk}
    \label{fig:id-cp-risk}
\end{subfigure}
\quad
\begin{subfigure}{0.48\textwidth}
\centering
	\includegraphics[width=\textwidth]{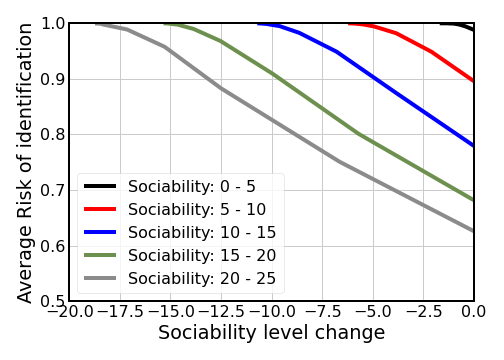}
    \caption{Identification risk versus sociability change}
    \label{fig:id-cp-change}
\end{subfigure}
\caption{Figure \ref{fig:id-cp-risk} shows the identification risk versus the RSSI threshold, while Figure \ref{fig:id-cp-change} shows the same metric with respect to the sociability change due to a varying RSSI threshold.}
\label{fig:cp-risk}
\end{figure*}

Figure \ref{fig:cp-risk} shows the identification risk for an individual when changing the RSSI threshold used to trace contacts. The rationale is that a lower threshold level increases the sociability of the individual, at the cost of possibly raising the number of false positives, while a higher threshold traces only close contacts, but decreasing the sociability level of the individual, hence making it more prone to the memory based attack.

Figure \ref{fig:id-cp-risk} shows the identification risk for users with different sociability levels when varying the RSSI threshold. Again it is possible to see that users which are more social are more protect against this kind of attack. However when we increase the RSSI threshold above \unit[-60]{dBm}, all users are exposed to the memory based identification attack regardless of their original sociability, since the set of user with which the graph is built is reduced due to the higher threshold set. This is also confirmed by Figure \ref{fig:id-cp-change}, where we show the same metric on the y axis, while the x axis represents the sociability change compared to an RSSI threshold set to \unit[-80]{dBm}. Again, when we raise the threshold, we reduce the number of users traced, while on the opposite reducing it increases the sociability level. We show that when reducing the sociability level of the individual due to a higher RSSI threshold, plotted on left side of the chart, the identification risk is higher compared to the right side of the chart, where we gradually achieve the sociability level obtained when using \unit[-80]{dBm} as RSSI threshold.

\begin{figure*}[t]
\centering
\begin{subfigure}{0.48\textwidth}
\centering
	\includegraphics[width=\textwidth]{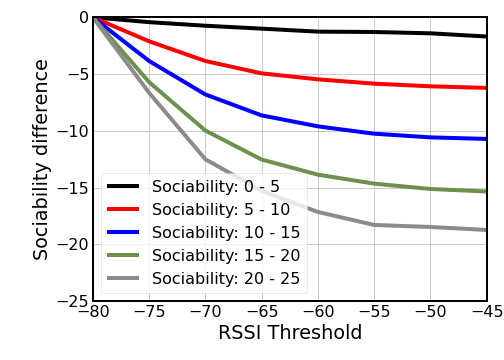}
    \caption{Sociability difference}
    \label{fig:soc-cp-change}
\end{subfigure}
\quad
\begin{subfigure}{0.48\textwidth}
\centering
	\includegraphics[width=\textwidth]{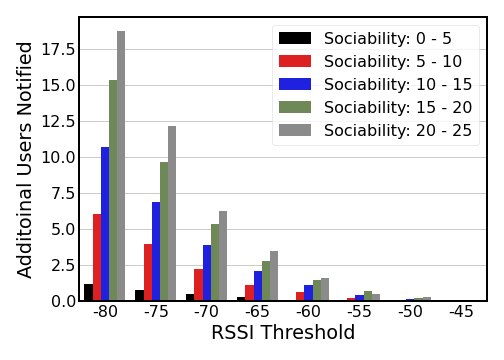}
    \caption{Additional users notified}
    \label{fig:add-cp}
\end{subfigure}
\caption{Figure \ref{fig:soc-cp-change} shows the identification risk versus the RSSI threshold, while Figure \ref{fig:add-cp} shows the same metric with respect to the sociability change due to a varying RSSI threshold.}
\label{fig:cp-soc-add}
\end{figure*}
We also show how the sociability of the individuals change due to a varying RSSI threshold, and what this would mean with respect to additional notifications sent to the users.

Figure \ref{fig:soc-cp-change} shows how the sociability of the individuals change when considering the lowest threshold we used in our experiments, which is \unit[-80]{dBm}. Clearly the sociability of users decreases as the RSSI threshold is raised. However, we can see that regardless of the sociability of the individuals, beyond \unit[-60]{dBm} no significant change is observed. Moreover, for low sociability individuals a change in the RSSI threshold increases the sociability less compared to other users. This is also confirmed by Figure \ref{fig:add-cp}, where we show how many more users would be notified due to the lower RSSI threshold compared to the highest threshold we tested. We can see that only highly social individual would notify a considerable higher amount of users, while low sociability individual (i.e. those with a sociability of 0-5) would notify less than 1 user more, but increasing their sociability also raises their chances to be protected against the memory based identification attack, as we have shown in previous figures.
\section{Conclusion and Discussion}
\label{sec:conclusion}
In this paper we have shown the memory based identification attack, a novel privacy issue which can be performed on de-centralised mobile contact tracing applications. We have shown how it is possible to build a graph leveraging the available information obtained by the BLE codes sent by users, and how this can be used to infer the positivity or negativity of the individuals. 

We have also discussed some techniques which can be used to mitigate the possibility for this attack to happen. All of these techniques can be implemented still keeping a de-centralised architecture: the \texttt{Variable time window} technique can be implemented in the server, with the MCTA providing feedback on whether more data is needed to perform the match; the \texttt{Contact injection} can be again performed on the server, though we showed that it is effective only when injecting real positive codes and not false ones, and we discussed the trade off between information freshness and privacy protection; finally the \texttt{Variable Contact Tracing Range} aims at raising the sociability level of the individual, and it can be implemented directly on the MCTA, possibly accounting for a reduced probability of infection if user were farther away, but still better protecting the user against the memory based identification attack.

Our results on two different real world datasets confirmed that the memory based identification attack can be indeed performed using solely the information available in the de-centralised MCTA system, and it can lead to the identification of the individuals who are positive or negative in the network. While understanding whether a user is negative may not have an impact, identifying a positive user that might be responsible for the infection may induce violence and self-justification of the notified person. The bottom line is -- this continuing practice, without consideration of the risks from memory-based re-identification, could potentially violate the privacy of individuals in the traces of decentralised MCTA systems.

Future work on this topic include personalized models dependent on the strength of ties of individuals, and the mobility patterns individuals perform. This would enable a more tailored modeling of each individual risk of being identified with a memory based attack. Moreover since our memory modeling is based on works which studied how people remember things, we also plan to collect real data which show to what extent people can remember encounters.

  \bibliographystyle{ACM-Reference-Format}
  \bibliography{citation}

\end{document}